\documentclass[conference]{IEEEtran}

\usepackage{amssymb}
\usepackage{amsmath}
\usepackage{cite}
\usepackage{amsfonts}
\usepackage{array}
\usepackage{subfigure}
\usepackage{color}
\usepackage{float}
\usepackage{stfloats}
\usepackage{psfrag}
\usepackage[ansinew]{inputenc}
\usepackage{arydshln}
\usepackage{amscd}
\usepackage{ntheorem}
\usepackage{graphicx}
\usepackage{amsopn}
\usepackage{indentfirst}

\begin{document}

\title{Amplitude Space Sharing among the Macro-Cell and Small-Cell Users}

\author{\authorblockN{Yafei Tian$^{\dag}$, Songtao Lu$^\ddag$, and Chenyang Yang$^\dag$}
\authorblockA{$^\dag$School of Electronics and Information Engineering, Beihang University, Beijing 100191, P. R. China\\
$^\ddag$Department of Electrical and Computer Engineering, Iowa
State University, Ames, IA 50011, USA\\
Emails: ytian@buaa.edu.cn, songtao@iastate.edu, cyyang@buaa.edu.cn}}

\maketitle

\begin{abstract}
The crushing demand for wireless data services will soon exceed the
capability of the current homogeneous cellular architecture. An
emerging solution is to overlay small-cell networks with the
macro-cell networks. In this paper, we propose an amplitude space
sharing (ASS) method among the macro-cell user and small-cell users.
By transmit layer design and data-rate optimization, the signals and
interferences are promised to be separable at each receiver and the
network sum-rate is maximized. The Han-Koboyashi coding is employed
and optimal power allocation is derived for the one small-cell
scenario, and a simple ASS transmission scheme is developed for the
multiple small-cells scenarios. Simulation results show great
superiority over other interference management schemes.
\end{abstract}

\begin{keywords}
Amplitude space sharing, Han-Koboyashi coding, interference channel,
power allocation, small-cell network.
\end{keywords}

\setlength \arraycolsep{1pt}
\section{Introduction}
With penetration of smartphones and tablets, the crushing demands of
wireless data traffic will soon exceed the capability of current
homogeneous cellular architecture. To deliver high-speed
transmission and consistent coverage, the multi-tier heterogeneous
architecture is emerging as a promising and economically sustainable
solution \cite{Yeh11,Hoydis11}. While the macro-cell base station
(MBS) provides near-universal coverage and supports fast mobility,
the low power small-cell access points (SAPs) provide high-capacity
transmission for hotspot zones. Generally, small-cells include
technologies variously described as femto-cells, pico-cells,
micro-cells and metro-cells \cite{SCF12}. By shrinking the
transmission range and intensifying spatial reuse of the spectrum,
these heterogeneous infrastructure elements can achieve significant
areal capacity gain.

Operated in licensed spectrum, small-cells introduce new forms of
interference to the macro-cell \cite{Lopez-Perez11}. For example,
femto-cells are typically deployed in ad hoc manner by costumers,
where there is not any constraint on node positions, and they might
be activated at any time. The large difference in transmission power
between MBS and SAP cause asymmetric interference modes both in
downlink and uplink. In some cases, we can control the mutual
interference by allocate appropriate powers to MBS and SAPs; but for
other cases the strong interference is inevitable, such as when the
macro-cell user (MUE) is located in the macro-cell edge and the
small-cell user (SUE) is located in the macro-cell center, the SUE
will receive dominant interference from the MBS. Fortunately, in the
latter scenarios, the SUE can still work effectively in an underlay
mode by the technique of interference cancelation.

Amplitude space sharing (ASS) means that we proactively design the
transmit layers and allocate data rates of each layer in the
network, so that at each receiver the interference can either be
treated as noise or be decoded and canceled. Compared with the
passive interference cancelation techniques \cite{Boudreau09}, ASS
creates the opportunities for interference cancelation and optimizes
the occupied spaces of each user for maximizing the multiuser
sum-rate. In the heterogeneous networks, although the MBS transmits
with large power, its data rate may not be high if the MUE is in the
cell edge; thus it has large potential for small-cell users to
utilize the residual amplitude space. Of course, the signals and
interferences can also be separated through transmit or receive
beamforming by employing multiple antennas
\cite{Chatzinotas12,Jeong11}, but the amplitude space and beam space
are two kinds of complementary degrees of freedom, we concentrate on
the studies of amplitude space in this paper.

With one small-cell coexisted with the macro-cell, the simultaneous
communications of MUE and SUE form a basic two-user interference
channel problem. The best known achievable scheme for this problem
is Han-Kobayashi (H-K) coding \cite{Han81}, where each user divide
its transmit information into private and common portions; the
private information is only decoded in the intended receiver and the
common information is decoded at both receivers. H-K coding is a
sophisticated ASS method because it divide each user's signal into
two layers and each layer has its transmit power and data rate. At
the receiver, totally four layers of signals and interferences will
share the amplitude space. To maximize the sum-rate of two users, we
develop an unified optimization framework that not only derives
classic results such as the sum-capacity in strong interference, but
also obtains the best known achievable sum-rate in weak
interference.

In multiple small-cells scenarios, the two-user H-K coding is not
applicable. In this case, we will simplify the transmitter-side
coding to just use one layer, but keep the receiver-side ASS as its
various possible forms. That means, at one SUE, the interference
from MBS may occupy the upper layer space; but at another SUE, the
interference from MBS may occupy the lower layer space. We will
allocate the transmit data rates of each user by a systematic method
inspired from the two-user H-K coding.

The rest of this paper is organized as follows. In Section II, we
derive the optimal power allocation of H-K coding for private and
common information in different interference scenarios. In Section
III, we propose the amplitude space sharing transmission scheme in
multiple small-cells environments. In Section IV, we will show the
connection between network geometry and interference mode, and
evaluate the performance gains of the ASS schemes over other
interference coordination schemes. Finally, Section V concludes the
paper.

\section{Optimized H-K Coding}
H-K coding defines private and common layers of transmission for
each user, but it does not specify the power allocation of each
layer. With different channel gains, we can optimize the power
allocation scheme to maximize the sum-rate of two users.

\subsection{Problem Formulation}
\begin{figure}[htp]
\begin{center}
\includegraphics[width=0.3\textwidth]{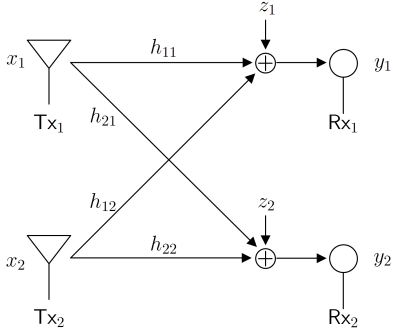}
\caption{Two user Gaussian interference channel
model.}\label{fig:channel_model}
\end{center}
\end{figure}

The nominal model of two-user Gaussian interference channel is shown
in Fig. \ref{fig:channel_model}, where $h_{ii}$ denotes the direct
signal link from $\text{Tx}_i$ to $\text{Rx}_i$ and $h_{ij}$ denotes
the cross interference link from $\text{Tx}_j$ to $\text{Rx}_i$, $
i,j\in \{1,2\}$. The transmit symbol of $\text{Tx}_i$ is $x_i$ that
is complex Gaussian and with power $P_i$, i.e., $x_i\in\mathbb{C},
\mathbb{E}\{|x_i|^2\} = P_i$. The received symbols of two users are
\begin{eqnarray}
y_1&=&h_{11}x_1+h_{12}x_2+z_1, \label{E:TxRx_model_1} \\
y_2&=&h_{21}x_1+h_{22}x_2+z_2, \label{E:TxRx_model_2}
\end{eqnarray}
where the noise $z_i\sim\mathcal{CN}(0,N_0)$ is circular symmetric
complex Gaussian with zero mean and variance $N_0$.

Define $\textsf{SNR}_1=|h_{11}|^2P_1/N_0$,
$\textsf{SNR}_2=|h_{22}|^2P_2/N_0$ as the receiver-side SNRs of user
1 and 2, and $\textsf{INR}_1=|h_{12}|^2P_2/N_0$,
$\textsf{INR}_2=|h_{21}|^2P_1/N_0$ as the receiver-side INRs of user
1 and 2, respectively. The private and common information are coded
separately and superimposed before transmission, i.e., $x_1 =
x_{1,p}+x_{1,c}$, $x_2=x_{2,p}+x_{2,c}$. The word `common' here does
not mean any data-sharing between two users.

For user $i$, $i=1,2$, assume that the power allocated to the
private information is $P_{i,p}$ and to the common information is
$P_{i,c}$, where $P_{i,p}+P_{i,c}=P_i$. The achievable data rates of
the private information and common information are denoted as
$R_{i,p}$ and $R_{i,c}$, respectively. At one receiver, the common
information from the intended user and interference user are first
decoded, and then the private information from the intended user is
decoded while regarding the remained private layer signal of the
interference user as background noise.

The achievable rates of the two private information are respectively
\begin{align}
&\;R_{1,p}=\log\left(1+\frac{|h_{11}|^2P_{1,p}}{|h_{12}|^2P_{2,p}+N_0}\right),\label{eq:R1_private}\\
&\;R_{2,p}=\log\left(1+\frac{|h_{22}|^2P_{2,p}}{|h_{21}|^2P_{1,p}+N_0}\right).\label{eq:R2_private}
\end{align}

For the common information, since the data from both transmitters
should be decodable at both receivers, the achievable rate region
are constrained by the intersection of the two multiple-access
channels capacity regions at $\text{Rx}_1$ and $\text{Rx}_2$.
Correspondingly, the achievable rates of two common information are
constrained as
\begin{eqnarray}
R^{(1)}_{1,c}&\leq\log\left(1+\frac{|h_{11}|^2P_{1,c}}{|h_{11}|^2P_{1,p}+|h_{12}|^2P_{2,p}+N_0}\right), \label{eq:R1_common_Rx1}\\
R^{(1)}_{2,c}&\leq\log\left(1+\frac{|h_{12}|^2P_{2,c}}{|h_{11}|^2P_{1,p}+|h_{12}|^2P_{2,p}+N_0}\right), \label{eq:R2_common_Rx1}\\
R^{(1)}_{1,c}+R^{(1)}_{2,c}&\leq\log\left(1+\frac{|h_{11}|^2P_{1,c}+|h_{12}|^2P_{2,c}}{|h_{11}|^2P_{1,p}+|h_{12}|^2P_{2,p}+N_0}\right),\label{eq:R_MAC1}\\
R^{(2)}_{1,c}&\leq\log\left(1+\frac{|h_{21}|^2P_{1,c}}{|h_{21}|^2P_{1,p}+|h_{22}|^2P_{2,p}+N_0}\right), \label{eq:R1_common_Rx2}\\
R^{(2)}_{2,c}&\leq\log\left(1+\frac{|h_{22}|^2P_{2,c}}{|h_{21}|^2P_{1,p}+|h_{22}|^2P_{2,p}+N_0}\right), \label{eq:R2_common_Rx2}\\
R^{(2)}_{1,c}+R^{(2)}_{2,c}&\leq\log\left(1+\frac{|h_{21}|^2P_{1,c}+|h_{22}|^2P_{2,c}}{|h_{21}|^2P_{1,p}+|h_{22}|^2P_{2,p}+N_0}\right),\label{eq:R_MAC2}
\end{eqnarray}
where (\ref{eq:R1_common_Rx1}), (\ref{eq:R2_common_Rx1}) and
(\ref{eq:R_MAC1}) construct the multiple-access channel capacity
region at $\text{Rx}_1$, and (\ref{eq:R1_common_Rx2}),
(\ref{eq:R2_common_Rx2}) and (\ref{eq:R_MAC2}) construct the
multiple-access channel capacity region at $\text{Rx}_2$. At both
receivers, the signals of private information serve as background
noise.

The optimization problem of power allocation to maximize the
sum-rate can be formulated as follows,
\begin{align}
\max_{P_{1,p}, P_{1,c}, P_{2,p}, P_{2,c}}&\;R_{1,p}+R_{1,c}+R_{2,p}+R_{2,c} \label{eq:sum_rate_opt}\\
\textrm{s.t.}\;&P_{1,p}+P_{1,c}=P_1,\nonumber \\
&\;P_{2,p}+P_{2,c}=P_2,\nonumber \\
&\;(\ref{eq:R1_private}), (\ref{eq:R2_private}),
(\ref{eq:R1_common_Rx1}), (\ref{eq:R2_common_Rx1}),
(\ref{eq:R_MAC1}), (\ref{eq:R1_common_Rx2}),
(\ref{eq:R2_common_Rx2}), (\ref{eq:R_MAC2}). \nonumber
\end{align}

\subsection{Optimization Results}
Consider sum-rate constraint of the two common messages, there are
totally four possibilities, i.e.,
\begin{eqnarray}\label{E:R_sum_common}
&& R_{\textsf{sum},c} = \nonumber \\
&& \min \left\{ R^{(1)}_{1,c}+R^{(1)}_{2,c},
R^{(2)}_{1,c}+R^{(2)}_{2,c}, R^{(1)}_{1,c}+R^{(2)}_{2,c},
R^{(2)}_{1,c}+R^{(1)}_{2,c} \right\}. \nonumber \\
\end{eqnarray}

\begin{table*}[!htp]
\caption{Summary of Optimal Power Allocation Schemes and Achievable
Sum-Rates}\label{table:summary}\centering \vskip 0in \small
\renewcommand{\arraystretch}{1.3}
\begin{tabular}{|c|c|c|c|c|}
\hline
\textbf{Interference Modes} & \textbf{Conditions} & $\boldsymbol{P_{1,p}}$ & $\boldsymbol{P_{2,p}}$ & $\boldsymbol{R_{\textsf{sum}}}$ \\
\hline Very Strong & $\begin{array}{l}
\textsf{SNR}_1<\frac{\textsf{INR}_2}{1+\textsf{SNR}_2} ~\text{and} \\
\textsf{SNR}_2 < \frac{\textsf{INR}_1}{1+\textsf{SNR}_1} \end{array}$ & $0$ & $0$ & $\log\left(1+\textsf{SNR}_1\right)+\log\left(1+\textsf{SNR}_2\right)$ \\
\hline
 Strong & $\begin{array}{l} \textsf{SNR}_1 <
\textsf{INR}_2 ~\text{and}\\ \textsf{SNR}_2 < \textsf{INR}_1
\end{array}$ & 0 & 0 & $\min \left\{ \begin{array}{c}
\log (1+\textsf{SNR}_2 + \textsf{INR}_2) \\
\log (1+\textsf{SNR}_1 + \textsf{INR}_1) \end{array} \right\}$  \\
\hline
Mixed 1 & $\begin{array}{l} \textsf{SNR}_1 \ge \textsf{INR}_2
~\text{and}\\ \textsf{SNR}_2 < \textsf{INR}_1 \end{array}$ & $P_1$ &
0 & $\min \left\{
\begin{array}{c}
\log (1+\textsf{SNR}_1 + \textsf{INR}_1) \\
\log (1+\textsf{SNR}_1) + \log \left(1+
\frac{\textsf{SNR}_2}{1+\textsf{INR}_2} \right)
\end{array} \right\}$ \\ \hline
Mixed 2 & $\begin{array}{l} \textsf{SNR}_1 < \textsf{INR}_2
~\text{and}\\ \textsf{SNR}_2 \ge \textsf{INR}_1 \end{array}$ & $0$ &
$P_2$ & $\min \left\{
\begin{array}{c}
\log (1+\textsf{SNR}_2 + \textsf{INR}_2) \\
\log \left(1+ \frac{\textsf{SNR}_1}{1+\textsf{INR}_1} \right) + \log
(1+\textsf{SNR}_2) \end{array} \right\}$  \\ \hline
Weak & $\begin{array}{l} \textsf{SNR}_1 \ge \textsf{INR}_2 ~\text{and}\\ \textsf{SNR}_2 \ge \textsf{INR}_1 \end{array}$ & $ P_{1,p}^*$ & $P_{2,p}^*$ & $ R_\textsf{sum}^* $  \\
\hline
Very Weak & $\gamma < 1$ & $P_1$ & $P_2$ &
$\log \left(1+ \frac{\textsf{SNR}_1}{1+\textsf{INR}_1} \right)+\log \left(1+ \frac{\textsf{SNR}_2}{1+\textsf{INR}_2} \right)$ \\
\hline
\end{tabular}
\vspace{8pt}
\begin{equation}\label{E:very_weak_cond}
\gamma =
\frac{\textsf{INR}_1\textsf{INR}_2\left(\textsf{SNR}_1\textsf{SNR}_2-\textsf{INR}_1\textsf{INR}_2+\textsf{SNR}_1-\textsf{INR}_2+\textsf{SNR}_2-\textsf{INR}_1\right)}{(\textsf{INR}_1-\textsf{SNR}_2)(\textsf{INR}_2-\textsf{SNR}_1)}.
\end{equation}
\begin{equation}\label{E:optimal_p1p}
\rho =
N_0\left(\frac{\sqrt{\frac{|h_{11}|^2|h_{22}|^2}{|h_{21}|^2|h_{12}|^2}(|h_{12}|^2-|h_{22}|^2)(|h_{21}|^2-|h_{11}|^2)\frac{1}{\alpha}}}{|h_{11}|^2|h_{22}|^2-|h_{21}|^2|h_{12}|^2}
-\frac{|h_{22}|^2-|h_{12}|^2}{|h_{11}|^2|h_{22}|^2-|h_{21}|^2|h_{12}|^2}\right).
\end{equation}
\hrulefill \vspace{-15pt}
\end{table*}



For each user the data rates of private information and common
information are competitive since their sum power is constrained.
Observing (\ref{eq:sum_rate_opt}) and (\ref{E:R_sum_common}), we
know that this is a max-min problem. The key idea in the solution
process is to use perspective transformation to optimize a
two-variable quadratic-fractional problem. Due to the limit space of
this paper, we do not provide detailed derivations here. The
optimized results of power allocation and achieved sum-rate are
listed in Table \ref{table:summary}.

According to the relationship of SNRs and INRs, we define six kinds
of interference channel modes as very strong, strong, mixed 1, mixed
2, weak, and very weak. The conditions of these interference modes
are also given in Table \ref{table:summary}. It should be noted that
$\textsf{SNR}_1$ and $\textsf{INR}_2$ depend on $P_1$, and
$\textsf{SNR}_2$ and $\textsf{INR}_1$ depend on $P_2$.

In weak interference mode, the optimal power allocations for the
private information of user 1 and user 2 have a linear relationship,
i.e.,
\begin{equation}\label{E:connect_p1p_p2p}
P_{2,p}^*=\alpha P_{1,p}^* + \beta,
\end{equation}
where
\[P_{1,p}^* = \max \{0, -\frac{\beta}{\alpha}, \rho \}, \]
\[\alpha=\frac{(|h_{11}|^2|h_{22}|^2-|h_{12}|^2|h_{21}|^2)P_2+(|h_{11}|^2-|h_{21}|^2)N_0}{(|h_{11}|^2|h_{22}|^2-|h_{12}|^2|h_{21}|^2)P_1+(|h_{22}|^2-|h_{12}|^2)N_0},\]
\[\beta=\frac{\left[(|h_{22}|^2-|h_{12}|^2)P_2+(|h_{21}|^2-|h_{11}|^2)P_1\right]N_0}{(|h_{11}|^2|h_{22}|^2-|h_{12}|^2|h_{21}|^2)P_1+(|h_{22}|^2-|h_{12}|^2)N_0},\]
and the expression of $\rho$ is given in (\ref{E:optimal_p1p}). The
achievable sum-rate in weak interference mode is
\begin{eqnarray}\label{E:R_sum_max_mw}
&&R_{\textsf{sum}}^*=\nonumber\\
&&~~\log\left(\frac{C_1C_2\frac{(|h_{11}|^2|h_{22}|^2-|h_{12}|^2|h_{21}|^2)P_{1,p}+(|h_{22}|^2-|h_{12}|^2)N_0}{(|h_{11}|^2|h_{22}|^2-|h_{21}|^2|h_{12}|^2)P_1+(|h_{22}|^2-|h_{12}|^2)N_0}}{(\alpha|h_{12}|^2P_{1,p}+\beta|h_{12}|^2+N_0)(|h_{21}|^2P_{1,p}+N_0)}\right).\nonumber\\
\end{eqnarray}

Observing from Table \ref{table:summary} we can find that, except in
weak interference mode, only one layer is required to achieve the
maximal sum-rate, either private or common. In very strong and
strong modes, each user only transmits common information; in mixed
mode, one user transmits common information and the other user
transmits private information; in very weak mode, each user only
transmits private information. In the general weak mode, both users
may transmit two layers of information.

Except in weak interference mode, the optimized sum-rates actually
achieve the sum-capacities. The capacity regions of two user
interference channel in very strong and strong modes are proved in
\cite{Carleial75} and \cite{Sato81}, respectively. The
sum-capacities in mixed and very weak modes are proved in
\cite{Shang09}. The sum-capacity in general weak mode is still an
open problem. To our knowledge, the result in Table
\ref{table:summary} is the best known achievable sum-rate in weak
interference mode.

\section{ASS by Multiple Small-Cell Users}
When multiple small-cells are deployed in a macro-cell, the
interference modes are more complicated. Theoretically, with $K$
small-cells coexisted with one macro-cell, it is a $(K+1)$-user
interference channel problem. The optimal transmission scheme design
is out of the scope of this paper. Instead, we use the insight
gained from the optimized H-K coding to design a simple ASS
transmission scheme for this kind of heterogeneous networks.

Suppose that $K$ small-cells are randomly deployed in the coverage
of the macro-cell, and each small-cell serves one SUE. Consider
downlink transmission, and the uplink case can be similarly
obtained. Given a position of the MUE, a two-user interference
channel problem is built between the MBS-MUE link and each SAP-SUE
link. Since the cross channel gains differ a lot, the constituted
link pairs might work in different interference modes.

Due to the low power of SAPs, the interference from other
small-cells are considered as background noise. We first express the
SNRs and INRs of each link pair as in the two-user interference
channel case, and then discuss how to design the transmit data rate
of each user so that the ASS scheme can work in the multiple
small-cells scenarios.

Define the SNR pairs and INR pairs of the MUE and the $k$-th SUE as
\begin{eqnarray}\label{E:K_pico_SNR_INR}
&&\textsf{SNR}_{{M},k} = \frac{|h_{00}|^2P_{M}}{\sum_{j=1,j\ne k}^K|h_{0j}|^2P_{{S},j}+N_0}, \nonumber \\
&&\textsf{INR}_{{M},k} = \frac{|h_{0k}|^2P_{{S},k}}{\sum_{j=1,j\ne k}^K|h_{0j}|^2P_{{S},j}+N_0}, \nonumber \\
&&\textsf{SNR}_{{S},k} = \frac{|h_{kk}|^2P_{{S},k}}{\sum_{j=1,j\ne k}^K|h_{kj}|^2P_{{S},j}+N_0}, \nonumber \\
&&\textsf{INR}_{{S},k} = \frac{|h_{k0}|^2P_{M}}{\sum_{j=1,j\ne
k}^K|h_{kj}|^2P_{{S},j}+N_0}, \nonumber
\end{eqnarray}
where $P_M$ is the transmit power of the MBS, $P_{S,j}$ is the
transmit power of the $j$-th SAP, $h_{k,j}$ is the channel gain from
the $j$-th SAP to the $k$-th SUE for $j,k \ne 0$, and $j=0$ denotes
the MBS and $k=0$ denotes the MUE.

If the MUE is located in the coverage of one small-cell, the link
pair may work in strong interference mode; if one SAP has a closer
distance to the MBS than the MUE, the link pair may work in mixed
interference mode; if another SAP has a longer distance to the MBS
than the MUE, the link pair may work in weak interference mode.
According to the optimized H-K coding, for each pair we can design a
transmission scheme that involves the optimal power and data rate
allocation for the layers of each user. However, the $K$ pairs
actually share the same MBS-MUE link, thus there is only one
possible transmission scheme for the MBS. For each layer of MBS's
transmit signal, we can only apply the lowest data rate of the $K$
calculated rates, otherwise there will be collision on the amplitude
space at some receivers.

For simplicity, every user only transmits one layer information, if
weak interference mode is encountered we use the transmission scheme
in very weak mode instead, i.e., treating the interference from the
other user as noise. With this assumption, there is no power
allocation any more and this single layer uses full transmit power.
However, whether this layer is private or common is not determined
at the transmitter, it is determined at each receiver. For the same
transmitted signal of MBS, it would behave as a common signal
(occupy the upper layer space) at the receiver of a cell-center SUE;
it would simultaneously behave as a private signal (occupy the lower
layer space) at the receiver of a cell-edge SUE.

In Table \ref{table:summary} we provide the achieved sum-rate, but
the data rate of each user is not specified. For different
interference modes, the sum-rates are obtained under different
multiple-access constraints as in (\ref{E:R_sum_common}). That means
for some cases the sum-rates are obtained on the corners of the rate
regions, for other cases the sum-rates are obtained on the side edge
of the rate regions.

For example, in the interference mode of mixed 1, if the first
sum-rate term in the minimization is satisfied, it is on the side
edge of the rate region and there are infinite combinations of the
data rates of two users; if the second sum-rate term is satisfied,
it is on the corner and there is only one possibility for the data
rates of two users.

For the latter case, the transmission rates of the $k$-th link pair
are calculated as
\[R_{M,k}=\log\left(1+\textsf{SNR}_{M,k}\right),~
 R_{S,k} =\log\left(1+\frac{\textsf{SNR}_{S,k}}{1+\textsf{INR}_{S,k}} \right). \]

For the former case, we choose to maximize $R_{M,k}$ since the real
transmission rate of MBS is the minimum of $K$ calculated values,
thus
\[R_{M,k}=\log\left(1+\textsf{SNR}_{M,k}\right),
 R_{S,k} =\log\left(1+\frac{\textsf{INR}_{M,k}}{1+\textsf{SNR}_{M,k}} \right). \]

After obtained $K$ pairs of data rates $\{R_{M,k}, R_{S,k}\}$, the
transmission rate of MBS is
\begin{equation}\label{E:K_pico_selection}
R_{M}=\min\{ R_{M,1}, R_{M,2}, \cdots, R_{M,K}\},
\end{equation}
and the throughput of the whole network is
\begin{equation}\label{E:K_pico_sum}
R_{\textsf{sum}} = R_{M} + \sum_{k=1}^K R_{S,k}.
\end{equation}

\section{Performance Evaluation}
In this section, we will apply the optimized H-K coding scheme to
the coordinated heterogeneous networks, and compare the network
throughput achieved by the ASS scheme with that of other schemes. We
will start from one small-cell scenario. The relationship between
the interference mode and network geometry is first shown, and then
the achievable sum-rates varying with the SAP positions are
demonstrated using the results of Table \ref{table:summary}.
Finally, we will examine the throughput of $K$ small-cells networks
along with the increasing of $K$.

Throughout this section, we consider some basic network
configurations as follows. The transmit power of the MBS is 46 dBm,
the transmit power of the SAP is 30 dBm, and the transmit power of
the user is 23 dBm. To show the performance gain brought purely by
the ASS scheme, single antenna is considered both in the BS side and
user side. The coverage of macro-cell is 500 m, where the measured
SNR at the cell edge is 5 dB. The radius of small-cell is set as 60
m. The path loss models for MBS and SAP are from 3GPP channel models
\cite{3GPP10}, i.e.,
\begin{eqnarray}\label{E:path_loss}
PL_{\;\text{MBS-UE}} &=& 15.3 + 37.6 \log_{10} (D), \nonumber\\
PL_{\;\text{SAP-UE}} &=& 30.6 + 36.7 \log_{10} (D), \nonumber
\end{eqnarray}
where $D$ is the distance between BSs and users,
$PL_{\text{MBS-UE}}$ applies to the path losses of MBS-MUE link and
MBS-SUE link, similarly $PL_{\text{SAP-UE}}$ applies to the path
losses of SAP-MUE link and SAP-SUE link. To avoid near-field effect,
SAP, SUE and MUE are not allowed to be close to the MBS within 35 m.

\subsection{Network Geometry and Interference Mode}
Given the position of MBS as in the center of a circular area, the
positions of MUE and SAP can be anywhere in the macro-cell, while
the SUE is located in the small-cell. The position distribution of
these nodes and their relative distances are called network
geometry.

The interference modes are determined by the relationship of SNRs
and INRs. In downlink, the values of $\textsf{SNR}_1$ and
$\textsf{INR}_2$ only depend on the channel gains $|h_{11}|$ and
$|h_{21}|$ since they have the same transmit power $P_1$. Similar
dependency happens to $\textsf{SNR}_2$ and $\textsf{INR}_1$.
According to the path loss models, the reference power and path loss
exponent are equal from one BS to different users, and the channel
gains are inversely proportional to the distances between each BS
and the two users. Thus the interference modes are determined by the
relative distances of direct and cross links, i.e.,
\begin{eqnarray}
&\text{Strong Interference:~~~} &~~D_{11} \ge D_{21}~~
\text{and}~~ D_{22} \ge D_{12},\nonumber \\
&\text{Mixed Interference 1:} &~~D_{11} < D_{21}~~
\text{and}~~ D_{22} \ge D_{12},\nonumber \\
&\text{Mixed Interference 2:} &~~D_{11} \ge D_{21}~~
\text{and}~~ D_{22} < D_{12},\nonumber \\
&\text{Weak Interference:~~~} &~~D_{11} < D_{21}~~ \text{and}~~
D_{22} < D_{12}.\nonumber
\end{eqnarray}
However, we cannot find a simple connection between the interference
mode and network geometry in the very strong and very weak modes.

In uplink, the users are transmitters and the BSs are receivers,
since the path losses from each user to different BSs subject to
different path loss formulas, there is no simple relationship as
well. We will illustrate the dependency through simulations.

\begin{figure}[htp!]
\begin{center}
\includegraphics[width=0.4\textwidth]{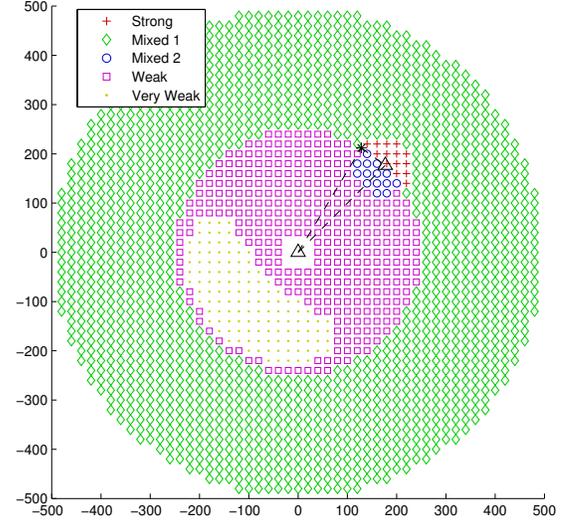}
\caption{The demonstration of various interference modes in
accordance with the locations of MUE, downlink. The central
`$\bigtriangleup$' denotes MBS, the upper right `$\bigtriangleup$'
denotes SAP, and the `$\ast$' denotes SUE.
}\label{fig:user_location_DL}
\end{center}
\end{figure}

\begin{figure}[htp!]
\begin{center}
\includegraphics[width=0.4\textwidth]{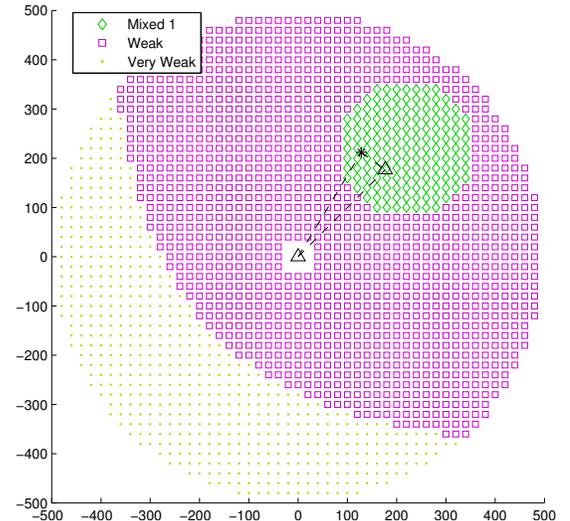}
\caption{The demonstration of various interference modes in
accordance with the locations of MUE, uplink. The central
`$\bigtriangleup$' denotes MBS, the upper right `$\bigtriangleup$'
denotes SAP, and the `$\ast$' denotes
SUE.}\label{fig:user_location_UL}
\end{center}
\end{figure}

Fix the positions of SAP, SUE, and change the position of MUE, we
can observe the changes of interference modes both in the downlink
and uplink transmissions. As shown in Fig.
\ref{fig:user_location_DL}, the position of MUE changes across the
macro-cell. We can see that totally five modes are appeared, but
most areas are belonged to the mixed, weak and very weak modes, the
strong mode only appears when the MUE is located in specific part of
the small-cell. Fig. \ref{fig:user_location_UL} illustrates the
interference modes in uplink as the MUE changes its position. We can
see that the relationship in uplink is quite different with that in
downlink scenarios.

\subsection{Comparisons with Other Schemes}
\begin{figure}[!t]
\begin{center}
\includegraphics[width=0.4\textwidth]{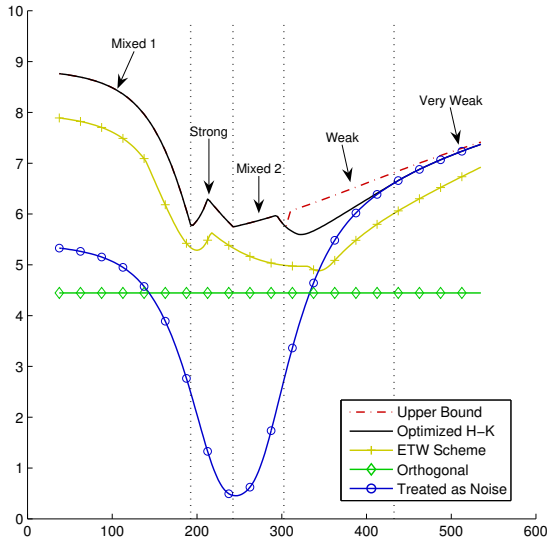}
\caption{The sum-rate of MUE and SUE when the SAP moves away from
cell center to cell edge while the SUE keeps relative position with
the SAP.} \label{fig:sum_rate_comparison}
\end{center}
\end{figure}

In last subsection, we fix the positions of SAP and SUE and change
the position of MUE across the whole macro-cell. Now, we fix the MUE
and move the SAP from the cell center to cell edge (35m - 500m),
while the SUE keeps relative position with the SAP, i.e., the SUE
moves along with the SAP. Different interference modes will be
encountered as the distance between SAP and MBS increases. The
achieved sum-rate is shown in Fig. \ref{fig:sum_rate_comparison},
where we also show the sum-capacity upper bound proved in
\cite{Etkin08}, the sum-rates of ETW's power allocation scheme
\cite{Etkin08}, the sum-rates of orthogonal transmission and
treating interference as noise. In \cite{Etkin08}, the private
information is allocated a power so that the INR of this layer
signal at the other receiver would equal to 1, or $\textsf{INR}_p =
1$, and it was proved that this scheme achieves the capacity region
outer bound to within 1 bit. The orthogonal transmission includes
the widely used fractional frequency reuse (FFR) scheme and the
almost blank subframe (ABSF) scheme, etc.

We can see that as the SAP moves from cell center to cell edge, the
link pair successively experience mixed 1, strong, mixed 2, weak and
very weak interference modes. In every mode, the optimized H-K
scheme performs the best comparing with other schemes, and achieves
the upper bound in strong and mixed modes. Although ETW's power
allocation scheme can achieve the capacity region outer bound to
within 1 bit, i.e., achieve the sum-capacity upper bound to within
two bits, the gap to the optimized power allocation scheme is
obvious. Orthogonal transmission has constant sum-rate as the SAP
moves, since the direct channel gains of the MBS-MUE link and the
SAP-SUE link do not change and there is no interference between
these two links. Treating interference as noise works better only in
very weak mode, and degrades seriously in other interference modes.

\begin{figure}[htp!]
\begin{center}
\includegraphics[width=0.4\textwidth]{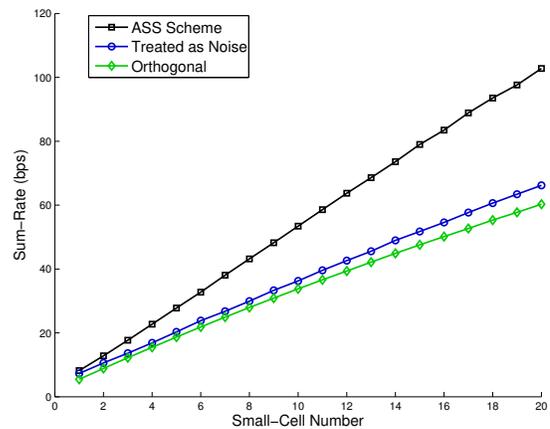}
\caption{The sum-rate of MUE and multiple SUEs when multiple SAPs
are randomly deployed in the macro-cell.}\label{fig:K_pico_d2}
\end{center}
\end{figure}

When $K$ small-cells are randomly deployed in the macro-cell, the
achieved sum-rate of the proposed ASS transmission scheme is shown
in Fig. \ref{fig:K_pico_d2}, where the achieved sum-rates of
orthogonal transmission and treating interference as noise are also
shown. In this simulation, we first set a virtual grid in the
macro-cell with separations of 120 m between the rows and columns,
then $K$ intersection points are randomly selected as the locations
of SAPs. In this manner, although randomly deployed, two SAPs keep a
minimum distance, and this is consistent with practical cellular
environments. The distance between MUE and MBS is fixed as $2/3$ of
the cell radius, and the SUE is randomly located in each small-cell.
From the figure we can see that, ASS scheme has great superiority
over the orthogonal transmission and treating interference as noise,
the sum-rate increasing slope almost doubles compared with the other
two schemes. Note that in multiple small-cells scenarios, we only
divide two time slots or frequency bands for orthogonal
transmission, the MBS uses one slot/band and all the SAPs use
another slot/band.


\section{Conclusion}
The overly of macro-cell with multiple small-cells can achieve
significant areal capacity gain. In this paper, we propose an
amplitude space sharing idea to manage the inter-cell interference.
Through the optimization of transmission powers and rates, at each
receiver the interference is promised to be separable and the
network sum-rate is maximized. In one small-cell scenarios, we
derived the optimal power allocation for H-K coding in different
interference modes. In multiple small-cells scenarios, we developed
a simple ASS transmission scheme which will double the network
throughput than the time/frequency orthogonal transmissions. The
principle of ASS can be easily applied in multiple-carrier and
multiple-antenna systems.

\bibliographystyle{IEEEtran}
\bibliography{IEEEabrv,my_ref}

\begin{thebibliography}{10}
\providecommand{\url}[1]{#1}
\csname url@rmstyle\endcsname
\providecommand{\newblock}{\relax}
\providecommand{\bibinfo}[2]{#2}
\providecommand\BIBentrySTDinterwordspacing{\spaceskip=0pt\relax}
\providecommand\BIBentryALTinterwordstretchfactor{4}
\providecommand\BIBentryALTinterwordspacing{\spaceskip=\fontdimen2\font plus
\BIBentryALTinterwordstretchfactor\fontdimen3\font minus
  \fontdimen4\font\relax}
\providecommand\BIBforeignlanguage[2]{{%
\expandafter\ifx\csname l@#1\endcsname\relax
\typeout{** WARNING: IEEEtran.bst: No hyphenation pattern has been}%
\typeout{** loaded for the language `#1'. Using the pattern for}%
\typeout{** the default language instead.}%
\else
\language=\csname l@#1\endcsname
\fi
#2}}

\bibitem{Yeh11}
S.-P. Yeh, S.~Talwar, G.~Wu, N.~Himayat, and K.~Johnsson, ``Capacity and
  coverage enhancement in heterogeneous networks,'' \emph{{IEEE} Wireless
  Commun. Mag.}, vol.~18, no.~3, pp. 32--38, June 2011.

\bibitem{Hoydis11}
J.~Hoydis, M.~Kobayashi, and M.~Debbah, ``Green small-cell networks,''
  \emph{{IEEE} Veh. Technol. Mag.}, vol.~6, no.~1, pp. 37--43, 2011.

\bibitem{SCF12}
S.~C. F.~W. Paper, ``Small cells - {W}hat's the big idea?'' Feb. 2012.

\bibitem{Lopez-Perez11}
D.~Lopez-Perez, I.~Guvenc, G.~{de la Roche}, M.~Kountouris, T.~Q.~S. Quek, and
  J.~Zhang, ``Enhanced intercell interference coordination challenges in
  heterogeneous networks,'' \emph{{IEEE} Wireless Commun. Mag.}, vol.~18,
  no.~3, pp. 22--30, June 2011.

\bibitem{Boudreau09}
G.~Boudreau, J.~Panicker, N.~Guo, and {etc.}, ``Interference coordination and
  cancellation for {4G} networks,'' \emph{{IEEE} Commun. Mag.}, vol.~47, pp.
  74--81, Apr. 2009.

\bibitem{Chatzinotas12}
S.~Chatzinotas and B.~Ottersten, ``Cognitive interference alignment between
  small cells and a macrocell,'' in \emph{ICT 2012}.

\bibitem{Jeong11}
Y.~Jeong, H.~Shin, and M.~Z. Win, ``Interference rejection combining in
  two-tier femtocell networks,'' in \emph{PIMRC 2011}.

\bibitem{Han81}
T.~S. Han and K.~Kobayashi, ``A new achievable rate region for the interference
  channel,'' \emph{{IEEE} Trans. Inf. Theory}, vol.~27, no.~1, pp. 49--60, Jan.
  1981.

\bibitem{Carleial75}
A.~B. Carleial, ``A case where interference does not reduce capacity,''
  \emph{{IEEE} Trans. Inf. Theory}, vol.~21, no.~5, pp. 569--570, Sept. 1975.

\bibitem{Sato81}
H.~Sato, ``The capacity of the {Gaussian} interference channel under strong
  interference,'' \emph{{IEEE} Trans. Inf. Theory}, vol.~27, no.~6, pp.
  786--788, Nov. 1981.

\bibitem{Shang09}
X.~Shang, G.~Kramer, and B.~Chen, ``A new outer bound and the noisy
  interference sum-rate capacity for {Gaussian} interference channels,''
  \emph{{IEEE} Trans. Inf. Theory}, vol.~55, no.~2, pp. 689--699, Feb. 2009.

\bibitem{3GPP10}
3GPP, ``Further advancements for {E-UTRA} physical layer aspects,'' in \emph{TR
  36.814 V9.0.0}, Mar. 2010.

\bibitem{Etkin08}
R.~H. Etkin, D.~N.~C. Tse, and H.~Wang, ``Gaussian interference channel
  capacity to within one bit,'' \emph{{IEEE} Trans. Inf. Theory}, vol.~54,
  no.~12, pp. 5534--5562, Dec. 2008.

\end{thebibliography}

\end{document}